\documentclass[useAMS,usenatbib]{mn2e}
\usepackage[dvips]{graphicx}
\usepackage{amsmath}
\usepackage{hyperref}
\usepackage{xcolor}
\usepackage{caption}
\usepackage{tablefootnote}
\usepackage{subcaption}
\newcommand{\bz}{$\langle B_z \rangle$}

\newcommand{\vsini}{$v \sin i$}
\newcommand{\kms}{km\,s$^{-1}$}

\newcommand{\msun}{$M_\odot$}

\newcommand{\mnras}{$MNRAS$}
\newcommand{\apj}{$ApJ$}
\newcommand{\apjl}{$ApJL$}
\newcommand{\aj}{$AJ$}
\newcommand{\aap}{A\&A}

\newcommand{\aaps}{A\&AS}

\newcommand{\pasp}{PASP}
\newcommand{\apjs}{ApJS}

\newcommand{\apss}{ApSS}

\newcommand{\nat}{Nature}

\newcommand{\solphys}{Sol.\ Phys.}

\title[Period evolution of HD\,142990]{The accelerating rotation of the magnetic He-weak star HD\,142990}
\author[M. Shultz]
{M.\ Shultz$^{1}$\thanks{E-mail: mshultz@udel.edu},
Th.\ Rivinius$^{2}$,
B.\ Das$^3$,
G.\ A. Wade$^{4}$,
P.\ Chandra$^3$\\
$^1$Annie Jump Cannon Fellow, Department of Physics and Astronomy, University of Delaware, 217 Sharp Lab, Newark, Delaware, 19716, USA\\
$^2$ESO - European Organisation for Astronomical Research in the Southern Hemisphere, Casilla 19001, Santiago 19, Chile\\
$^3$1National Centre for Radio Astrophysics, Tata Institute of Fundamental Research, Pune University Campus, Pune-411007, India\\
$^4$Department of Physics and Space Science, Royal Military College of Canada, Kingston, Ontario K7K 7B4, Canada\\
}
\begin{document}

\date{}

\pagerange{\pageref{firstpage}--\pageref{lastpage}} \pubyear{2018}

\maketitle

\label{firstpage}

\begin{abstract}
HD\,142990 (V\,913\,Sco; B5\,V) is a He-weak star with a strong surface magnetic field and a short rotation period ($P_{\rm rot} \sim 1$~d). While it is clearly a rapid rotator, recent determinations of $P_{\rm rot}$ are in formal disagreement. In this paper we collect magnetic and photometric data with a combined 40-year baseline in order to re-evaluate $P_{\rm rot}$ and examine its stability. Both period analysis of individual datasets and $O-C$ analysis of the photometric data demonstrate that $P_{\rm rot}$ has decreased over the past 30 years, violating expectations from magnetospheric braking models, but consistent with behaviour reported for 2 other hot, rapidly rotating magnetic stars, CU\,Vir and HD\,37776. The available magnetic and photometric time series for HD\,142990 can be coherently phased assuming a spin-up rate $\dot{P}$ of approximately $-0.6$ s/yr, although there is some indication that $\dot{P}$ may have slowed in recent years, possibly indicating an irregular or cyclic rotational evolution.
\end{abstract}


\section{Introduction}

Of order 10\% of OBA stars possess detectable surface magnetic fields \citep[e.g.][]{2017MNRAS.465.2432G,2019MNRAS.483.2300S} which are generally topologically simple \citep[approximately dipolar;][]{1987ApJ...323..325B,2018MNRAS.475.5144S,2019MNRAS.483.3127S}, strong \citep[with a surface strength at the magnetic pole above 300 G;][]{2007A&A...475.1053A,2019MNRAS.483.3127S}, and demonstrate no detectable intrinsic evolution over observational timescales \citep[e.g.][]{2018MNRAS.475.5144S} but gradual weakening over evolutionary timescales \citep[e.g.][]{land2007,land2008,2016A&A...592A..84F}. These properties have led to their characterization as `fossil' magnetic fields \citep[e.g.][]{2004Natur.431..819B,2015IAUS..305...61N}, i.e.\ magnetic flux generated at an earlier evolutionary stage and preserved via persistent magneto-hydrostatic equilibria.  The fossil fields of massive stars are thus fundamentally different from the magnetic fields of cool stars, which are maintained by contemporaneous dynamos.

Magnetic hot stars can be observationally distinguished from their non-magnetic kin by several phenomena which are empirically known to co-occur with the presence of a magnetic field. The best-known diagnostics, valid for stars later than about B1, are atmospheric chemical peculiarities: significant photospheric over- or under-abundances of He, Fe, Si, as well as rare-earth elements \citep[e.g.][]{2010AA...524A..66S,2013AA...551A..30B,2014A&A...565A..83K,2015MNRAS.447.1418Y,2015MNRAS.449.3945S,2017A&A...605A..13K,2019MNRAS.483.2300S}. These are believed to arise due to radiative diffusion in atmospheres stabilized by strong magnetic fields \citep[e.g.][]{michaud1970,michaud1981,2015MNRAS.454.3143A}. These chemical peculiarites typically exhibit nonuniformities of up to several dex across the stellar surface, leading to photometric and spectroscopic variability modulated according to the star's rotation \citep[e.g.][]{2009A&A...499..567K,2015A&A...576A..82K}. This variability is strictly periodic, allowing rotational periods to be easily determined from the spectroscopic and/or photometric variations of magnetic Chemically Peculiar (mCP) stars. 

Another distinguishing feature is that magnetic hot stars are, as a population, more slowly rotating than non-magnetic stars \citep[e.g.][]{2018MNRAS.475.5144S}. This is consistent with the expectation that angular momentum is efficiently lost through the magnetically confined stellar wind \citep[][]{wd1967,ud2009}. Period evolution has been directly measured for three magnetic hot stars: $\sigma$ Ori E, CU\,Vir, and HD\,37776. All are stars with relatively short rotational periods, of the order of one day, for which extensive and well-sampled datasets spanning decades are available. A spindown rate qualitatively compatible with predictions from magnetospheric braking models was reported by \cite{town2010} for the magnetic B2 star $\sigma$ Ori E. Curiously, the rotational period of the B6 star CU Vir has been observed to decrease with time, i.e.\ the star's rotation has {\em accelerated} \citep{1998A&A...339..822P,miku2011}. Spin-up was also reported by \cite{miku2011} for the magnetic B2 star HD\,37776, a suspicion later confirmed by \cite{2016CoSka..46...95M} with additional observations. 

HD\,142990 (V\,913\,Sco; B5\,V) is a mCP He-weak star \citep{1974A&A....36...57N} distinguished by rapid rotation \citep[$P_{\rm rot} \sim 0.98$~d; ][]{1996A&A...311..230C,2018MNRAS.475.5144S,2018A&A...616A..77B} and a strong \citep[several kG;][]{2016PhDT.......390S} surface magnetic field. The star shows extremely weak ultraviolet and H$\alpha$ emission originating in its circumstellar magnetosphere \citep{2004A&A...421..203S}. It was recently suggested by \cite{2018MNRAS.478.2835L}, based on a survey at 200 MHz, to display coherent, highly polarized radio emission consistent with electron-cyclotron maser emission. 


Rotational periods have recently been published by \cite{2018MNRAS.475.5144S}, who used magnetic measurements with a long temporal baseline, and \cite{2018A&A...616A..77B}, who used Kepler-2 photometry with a much shorter temporal baseline but much higher precision. The two periods are similar, but are in formal disagreement. This motivates the re-examination of the star's rotational period, which is the subject of this paper. We review the published photometric and magnetic time series in \S~\ref{sec:obs}, along with previously unpublished SMEI photometry. In \S~\ref{sec:results} we show that the rotational period has almost certainly decreased over the 30-year time-span of observation, and in \S~\ref{sec:conclusion} we discuss the implications of this result. 

\section{Observations}\label{sec:obs}

\begin{table}
\begin{minipage}{8.5cm}
\caption{Summary of available datasets. Columns indicate the name of the dataset; the type of data, either P(hotometric) or M(agnetic); the year (or mean year) of the dataset's acquisiion; the timespan of dataset; and the original work in which is was published$^*$. Note that the STEREO data are not publicly available, but are included here for completeness.}
\label{obstab}

\resizebox{8.5 cm}{!}{
\begin{tabular}{l l l l l l}
\hline\hline
Dataset & Type & Year & Timespan & $N_{\rm obs}$ & Reference \\
\hline
Str\"omgren & P & 1976 & 8 d & 12 & $a$ \\ 
Photopolarimetric & M & 1980 & 4.1 yr & 14 & $b$ \\ 
Photopolarimetric & M & 1988 & 0.26 d & 4 & $c$ \\ 
Tycho & P & 1991 & 2.1 yr & 147 & $d$ \\
Hipparcos  & P & 1991 & 3.0 yr & 111 & $e$ \\ 
Str\"omgren & P & 1991 & 4 d & 18 & $f$ \\ 
Str\"omgren & P & 1993 & 2.2 yr & 144 & $g$ \\ 
SMEI & P & 2007 & 8.6 yr & 18851 & $h$ \\ 
STEREO & P & 2007 & 4.3 yr & 6000 & $i$ \\
K2 & P & 2014 & 77 d & 3293 & $j$ \\ 
ESPaDOnS & M & 2015 & 5.9 yr & 15 & $k$ \\
\hline\hline
\end{tabular}
}

$^*${\footnotesize {\em Reference key}: $a$: {\protect\cite{1977A&AS...30...11P}}; $b$: \protect\cite{1983ApJS...53..151B}, $c$: {\protect\cite{1993A&A...269..355B}}; $d$: \protect\cite{2000A&A...355L..27H}; $e$: \protect\cite{vanleeuwen2007}; $f$: {\protect\cite{1995A&AS..109..329M}}; $g$: {\protect\cite{1995A&AS..113...31S}}; $h$: This work; $i$: {\protect\cite{2012MNRAS.420..757W}}; $j$: {\protect\cite{2018A&A...616A..77B}}; $k$: \protect\cite{2018MNRAS.475.5144S}.}
\end{minipage}
\end{table}

The characteristics and origins of the datsets used in this work are summarized in Table \ref{obstab}. 

The most recent magnetic measurements of HD\,142990 were performed between 2011 and 2017 using ESPaDOnS spectropolarimetry obtained by the Magnetism in Massive Stars (MiMeS) large program at the Canada-France-Hawaii Telescope \citep{2016MNRAS.456....2W}, together with several observations obtained by PI programs\footnote{Program codes 14AC10 and 17AC16.}. The analysis of these data was described by \cite{2018MNRAS.475.5144S}. In addition to the modern data, we also compiled the photopolarimetric magnetic measurements reported by \cite{1983ApJS...53..151B} and \cite{1993A&A...269..355B}. These data were also analyzed by \cite{2018MNRAS.475.5144S}, however several of the measurements of \citeauthor{1983ApJS...53..151B} were overlooked in that analysis. The combined magnetic dataset spans 1978 to 2017, but mostly samples the beginning and end of this time-frame.

The spectropolarimetric \bz~measurements used here were performed using H lines, since these measurements should be unaffected by distortions introduced by surface chemical abundance inhomogeneities \citep{1979ApJ...228..809B,bl1980,2018MNRAS.475.5144S}. The photopolarimetric data were also obtained using H lines, in particular the wings of H$\beta$. The different measurement systems can in principle introduce systematic discrepancies between datasets. In practice, however, agreement between high-resolution H line \bz~measurements and photopolarimetric \bz~measurements is quite good, thus any such differences must be less than the intrinsic scatter in the latter \citep[e.g.][]{oks2012,2018MNRAS.475.5144S}.


Several photometric time series are also available, spanning the time frame from 1978 to 2014, but with more favourable sampling than the magnetic data. We acquired the ground-based Str\"omgren photometry reported by \cite{1977A&AS...30...11P} from the mCPod photometric database \citep{2007AN....328...10M}\footnote{Available at \url{http://mcpod.physics.muni.cz/}.} We also obtained Str\"omgren photometry from the catalogues published by \cite{1995A&AS..109..329M} and \cite{1995A&AS..113...31S} via VizieR  \citep{2000A&AS..143...23O}. These data were originally analyzed by \cite{1996A&A...311..230C}, together with their own data which, unfortunately, are not publicly available. We downloaded Hipparcos photometry \citep{perry1997,vanleeuwen2007} and Tycho $BV$ photometry \citep{2000A&A...355L..27H} from VizieR. We obtained the Kepler light curve from  the Mikulski Archive for Space Telescopes (MAST\footnote{Available at \url{https://mast.stsci.edu}.}). As there are numerous Kepler reductions available, we selected the best light curve by eye by phasing the data with the rotation period determined by \cite{2018A&A...616A..77B}, choosing the light curve file \verb k2sff203814494-c02_lc ~and the corrected flux in aperture 9.

We use data obtained by the Solar Mass Ejection Imager, SMEI \citep{2004SoPh..225..177J}. The data have been reprocessed by Jackson (priv. comm.) and are available on request. The new processing identifies which of the three cameras each individual datapoint was observed with. This allows for much better background and trend corrections, since these are individual to each camera. The computational methods for this are described by \citet{2016A&A...593A.106R}. In detail the dataset was split into three per-camera subsets. In each set, the variability is strongly dominated by annual and daily signals. To remove them, first the variability in the vicinity, meaning within typically 3\%, of 1 cycle/yr and the first nine harmonics were removed. In the next step the same was done for 1 cycle/d and its first four harmonics, but only removing signals within narrower windows of $\pm 0.2$\%. The resulting dataset was visually clipped for strong outliers, and the frequency removal procedure was repeated. The final datasets can then either be re-merged to analyse all cameras together, or further split to allow an analysis per camera and season, for example. It turns out that the data quality delivered by one camera degraded much more strongly over the years than that of the others, so data from this camera was included in the analysis only for the first few mission years.

We also explored the photometric time series acquired by the All Sky Automated Survey Supernovae \citep[ASAS-SN;][]{2017PASP..129j4502K,2018MNRAS.477.3145J} and by the Super-Wide Angle Search for Planets \citep[SuperWASP][]{2010A&A...520L..10B}. However, even after detrending and outlier-rejection, the precision of these datasets is insufficient to detect HD\,142990's photometric variability (Andzrej Pigulski, priv.\ comm.).


\section{Results}\label{sec:results}

\subsection{Previous period determinations}\label{subsec:published_periods}

   \begin{figure*}
   \centering
\begin{tabular}{ccc}
   \includegraphics[trim=30 30 30 30, width=5.5cm]{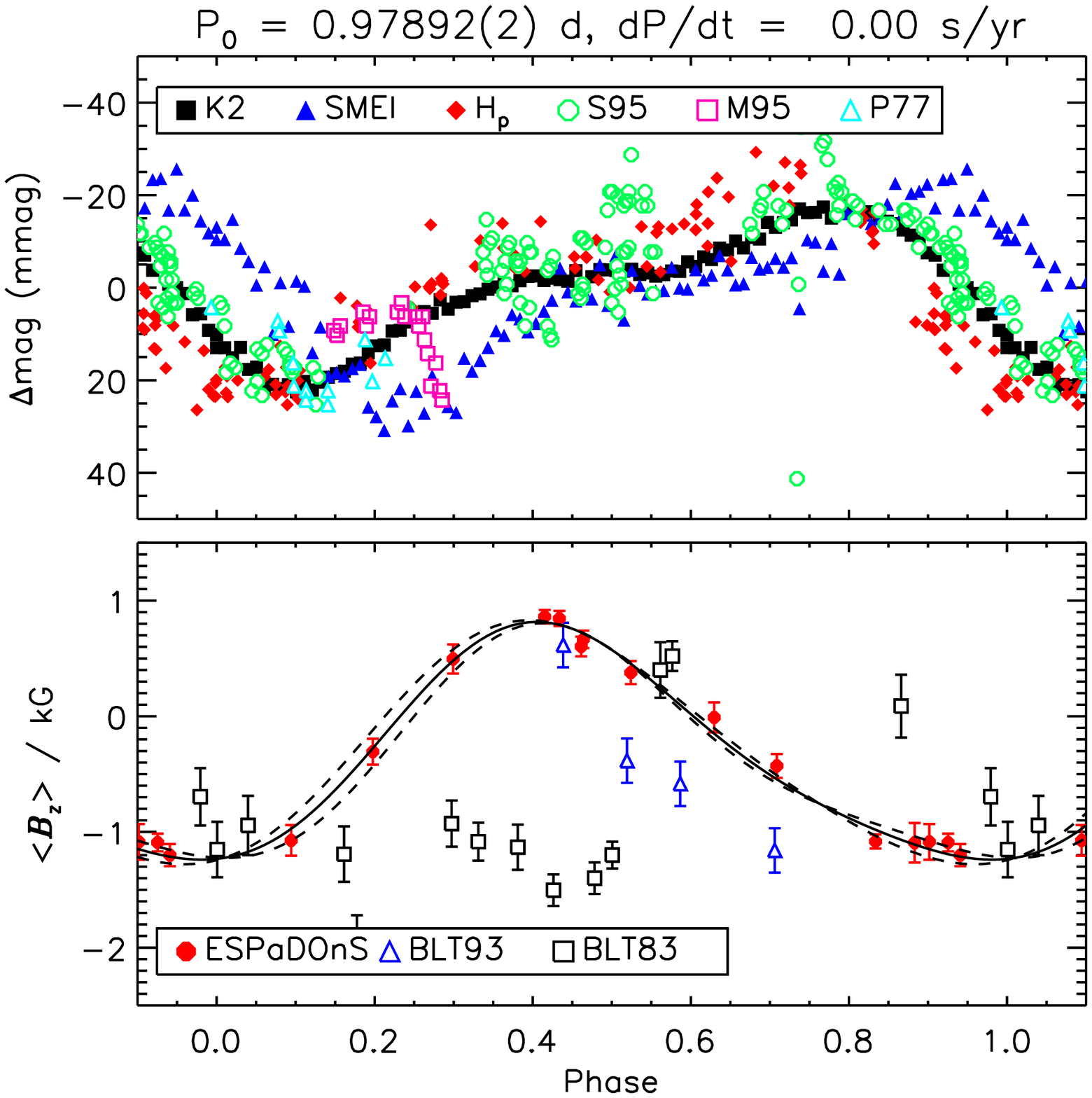} &
   \includegraphics[trim=30 30 30 30, width=5.5cm]{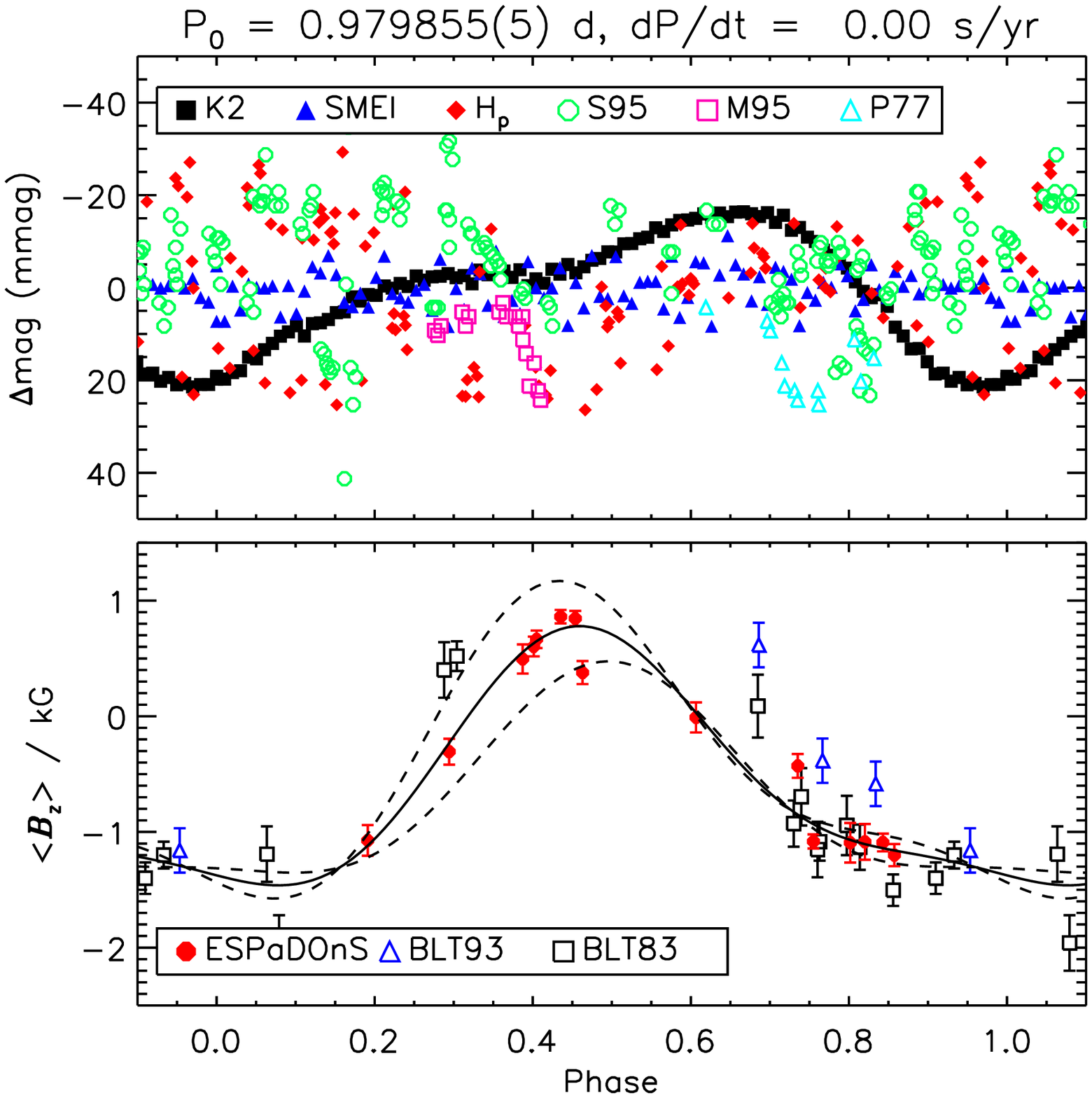} &
   \includegraphics[trim=30 30 30 30, width=5.5cm]{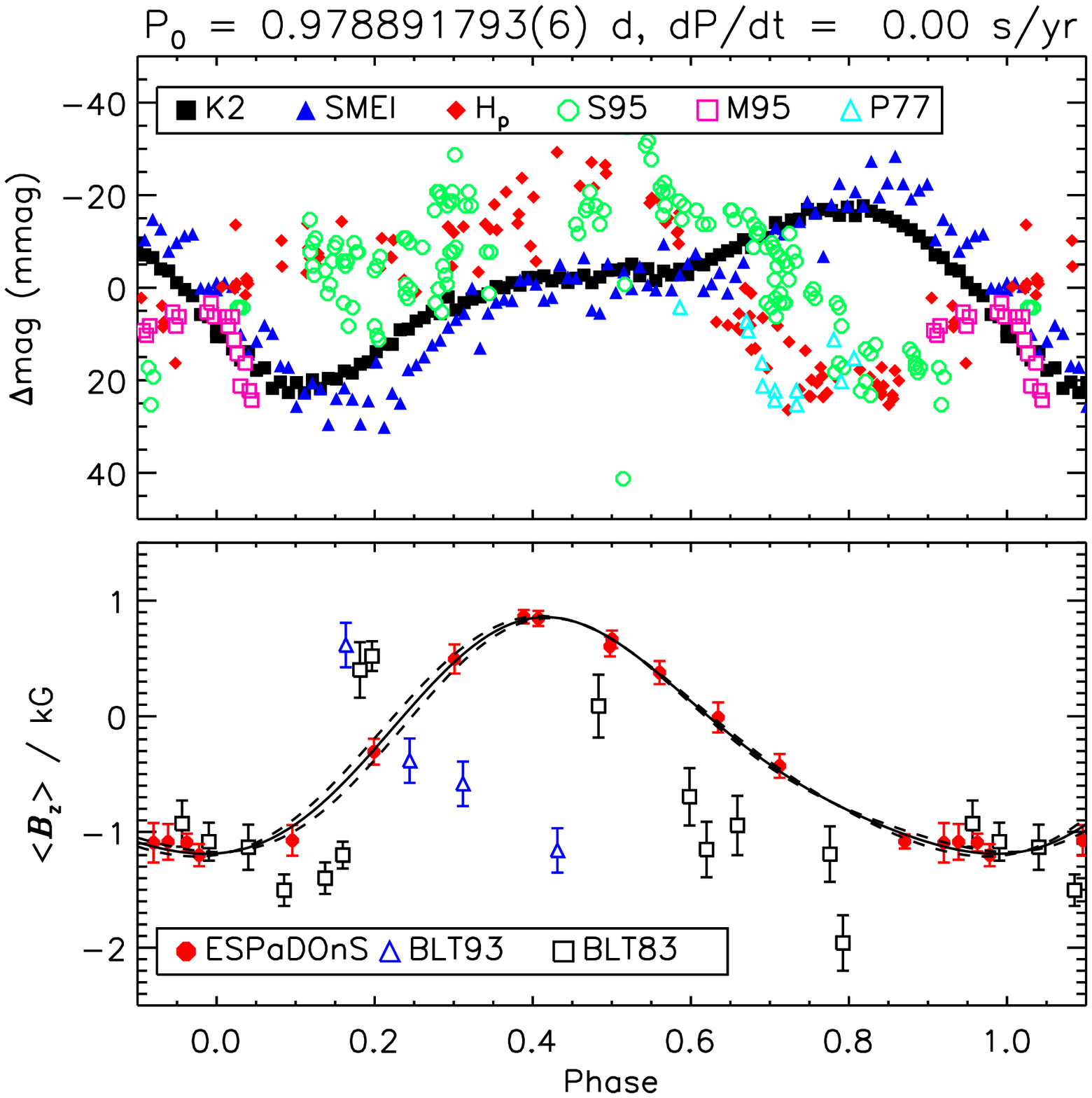} \\
\end{tabular}
      \caption[]{Photometric ({\em top}) and magnetic \bz~({\em bottom}) measurements, phased with periods determined from ({\rm left to right}) K2 photometry, all \bz~measurements, and all photometric measurements. \bz~measurements were obtained from ESPaDOnS by \protect\cite{2018MNRAS.475.5144S} and photopolarimetric data by \protect\citet[BLT83]{1983ApJS...53..151B} and \protect\citet[BLT93]{1993A&A...269..355B}. The solid and dashed curves show respectively the best second-order harmonic fit to the ESPaDOnS \bz~data and the 1$\sigma$ fit uncertainty. Photometric data are from K2, SMEI, Hipparcos ($H_{\rm p}$), and Str\"omgren $y$ photometry published by \protect\citet[S95]{1995A&AS..113...31S}, \protect\citet[M95]{1995A&AS..109..329M}, and \protect\citet[P77]{1977A&AS...30...11P}. For clarity the K2 and SMEI data have been binned by phase, using bin sizes of 0.01 cycles. The mean magnitude of each dataset was subtracted for display purposes.}
         \label{hd142990_bz_phot_rf_constp}
   \end{figure*}

\begin{table*}
\centering
\caption[]{Summary of periods from the literature and determined in the present work. The table is organized in order of the year corresponding to the mean observation time of the dataset(s) used to determine $P_{\rm rot}$. The second column gives the mean HJD of the datasets used for each period. The fourth column gives the difference in $P$ relative to the K2 period. The Dataset column indicates the data used to determine the periods, corresponding to the reference key in Table \ref{obstab}. Origin corresponds to the work in which the period was published. Periods in boldface were adopted for analysis.}
\label{prottab}
\begin{tabular}{l l l c l l}
\hline\hline
Year & HJD -   & $P_{\rm rot}$ & $\Delta P$ & Dataset & Origin \\
     & 2400000 & (d)           & (s)        &         &           \\
\hline
1976 & 42826 & 0.976(2) & $-252 \pm 172$  & $a$ & {\protect\cite{1977A&AS...30...11P}} \\
1981 & 44948 & 0.98292(2) & $345 \pm 2$ & $b,c$ & {\protect\cite{1993A&A...269..355B}} \\
1981 & 44948 & {\bf 0.97910(4)} & $15 \pm 3$ & $b,c$ & This work \\
1986 & 46774 & {\bf 0.97907(1)} & $13 \pm 1$ & $a,f,g$ & {\protect\cite{1996A&A...311..230C}} \\ 
1991 & 48307 & 0.97904(4) & $10 \pm 4$ & $e$ & {\protect\cite{2011MNRAS.414.2602D}} \\
1991 & 48307 & {\bf 0.97901(4)} & $8 \pm 4$ & $d,e$ & This work \\
1993 & 49155 & {\bf 0.97902(5)} & $9 \pm 4$ & $g$ & This work \\
1995 & 49896 & 0.978891793(6) & $ -2.437 \pm 0.0005 $ & $a,e,f,g,h,j$ & This work \\
1997 & 50529 & 0.978832(2) & $-7.6 \pm 0.2$ & $b,c,k$ & {\protect\cite{2018MNRAS.475.5144S}} \\
1997 & 50529 & 0.979855(5) & $80.8 \pm 0.4$ & $b,c,k$ & This work \\
2007 & 54245 & {\bf 0.97890(5)} & $-1 \pm 4$ & $h$ & This work \\
2007 & 54387 & {\bf 0.9789(1)} & $-2 \pm 8$ & $i$ & {\protect\cite{2012MNRAS.420..757W}} \\
2014 & 56933 & {\bf 0.97892(2)} & $0 \pm 2$ & $j$ & {\protect\cite{2018A&A...616A..77B}} \\
2015 & 57227 & {\bf 0.97887(6)} & $-4 \pm 5$ & $k$ & This work \\
\hline\hline
\end{tabular}
\end{table*}

Periods collected from the literature are summarized in Table \ref{prottab}. 

The most recent rotational period of HD\,142990 was determined by \cite{2018A&A...616A..77B}, who used Kepler 2 (K2)  photometry to determine $P_{\rm rot} = 0.97892(2)$~d. This period is close to, but not formally compatible with, the period given by \cite{2018MNRAS.475.5144S}, 0.978832(2)~d, which was obtained by combining longitudinal magnetic field \bz~measurements from high-resolution ESPaDOnS spectropolarimetry with photopolarimetric \bz~measurements presented by \cite{1983ApJS...53..151B} and \cite{1993A&A...269..355B}. \cite{2012MNRAS.420..757W} determined $P_{\rm rot} = 0.9789(1)$~d from STEREO photometry, which is compatible with either period within its large uncertainty. 

The earliest period, 0.976(2)~d, was provided by \cite{1977A&AS...30...11P} using ground-based Str\"omgren photometry. \cite{1996A&A...311..230C} combined their own measurements with those of \citeauthor{1977A&AS...30...11P}, \cite{1995A&AS..109..329M}, and \cite{1995A&AS..113...31S}, to find $P_{\rm rot} = 0.97907(1)$~d. Hipparcos photometry \citep{perry1997,vanleeuwen2007} was used to determine a period of $0.97904(4)$~d by \cite{2011MNRAS.414.2602D}. 

\cite{1993A&A...269..355B} found $P_{\rm rot } = 0.98292(2)$~d by combining all photopolarimetric \bz~measurements. This period is incompatible with all of the other periods.

The left panel of Fig.\ \ref{hd142990_bz_phot_rf_constp} shows all available photometric and magnetic data phased with the K2 period. The epoch was set at the time of minimum \bz~one cycle before the first ESPaDOnS observation, as determined via a harmonic fit. This period produces a reasonable phasing of the K2 and Str\"omgren photometry, but yields phase shifts of about 0.1 cycles relative to the the Hipparcos and SMEI data. While a reasonable harmonic fit to the ESPaDOnS data can be achieved using this period, it does not produce an acceptable phasing of all available \bz~measurements. 

All of the periods obtained since the year 2000 are approximately consistent with one another; the two periods determined using photometric data obtained prior to the year 2000 are also consistent with one another. At the same time, the difference between the \cite{2018A&A...616A..77B} and \cite{1996A&A...311..230C} periods, $0.00015$~d, is almost 8 times larger than the formal uncertainty in the \citeauthor{2018A&A...616A..77B} period (the less precise of the two). This suggests that the period may have changed. 

\subsection{Period analysis}\label{subsec:period_analysis}


We now turn to a re-examination of the period analysis of the individual datasets; the results are summarized in Table \ref{prottab}. Period analysis was performed using standard Lomb-Scargle statistics \citep{1976ApSS..39..447L, 1982ApJ...263..835S}, utilizing both the {\sc idl} program {\sc periodogram}\footnote{https://hesperia.gsfc.nasa.gov/ssw/gen/idl/\\util/periodogram.pro} and the {\sc period04} package \citep{2005CoAst.146...53L}. The uncertainties were determined in the same manner as by \cite{2018MNRAS.475.5144S}, i.e.\ according to the analytic method described by \cite{1976fats.book.....B}.


Examining the ESPaDOnS data in isolation yields $P_{\rm rot} = 0.97887(6)$~d, consistent within uncertainty with the K2 period \citep{2018A&A...616A..77B}. The period determined from all available magnetic data \citep[i.e.\ including the previously overlooked \bz~measurements of][]{1983ApJS...53..151B} is $0.979855(5)$~d. The photometric and magnetic data are shown phased with this period in the middle panel of Fig.\ \ref{hd142990_bz_phot_rf_constp}. Phasing \bz~with this period and fitting the ESPaDOnS data with a second-order sinusoid yields a reduced $\chi^2 = 5$, i.e.\ a poor fit. This period also does not provide a coherent phasing of the majority of the photometric dataset. Notably, the variation in the phase-binned SMEI data almost completely disappears, likely because observations are being binned as though they were at the same phase when, in fact, they are not. 

We can also combine all available photometric data (shifted to the mean magnitude of each dataset, as in Fig.\ \ref{hd142990_bz_phot_rf_constp}), which yields $P_{\rm rot} = 0.978891793(6)$. This makes the assumption that there are no large, systematic differences in light curve morphology between different bandpasses (this is explored in greater detail in \S~\ref{bandpass}). The left panel of Fig.\ \ref{hd142990_bz_phot_rf_constp} shows the datasets phased with this period. It achieves a reasonable phasing of the K2 and SMEI data, but a poor phasing of the other photometric datasets, and while it phases the ESPaDOnS data well, it does not coherently phase the modern and historical \bz~data.

If the period is changing, the accuracy of the \cite{1996A&A...311..230C} result is questionable as the dataset they used spans approximately 15 years. However, our own analysis of the \cite{1995A&AS..113...31S} photometry used by \citeauthor{1996A&A...311..230C}, which has a much more restricted time-frame (about 2 years), finds $P_{\rm rot} = 0.97902(5)$, which is compatible with the \citeauthor{1996A&A...311..230C} period. 

We analyed the Hipparcos data together with the Tycho $BV$ data, with the weighted mean period for the three datasets yielding $P_{\rm rot} = 0.97901(4)$~d, compatible with the \cite{2011MNRAS.414.2602D} result. This is also compatible with the period we obtained from \cite{1995A&AS..113...31S} Str\"omgren photometry, which was obtained at close to the same time.

We next turn to a re-examination of the period determined by \cite{1993A&A...269..355B}, which is incompatible with any of the other periods. This is a sparse dataset spanning a large time-frame (about a decade), and as a result the periodogram has numerous closely-spaced peaks that are likely to yield false positives. Limiting the period window to 3 times the range spanned by the \cite{1996A&A...311..230C} and ESPaDOnS periods (which bracket the lower and upper extremes of the derived periods), the most significant peak is at 0.97910(4)~d; this is within uncertainty of the \citeauthor{1996A&A...311..230C} period. A window of 3 times the range of the \citeauthor{1996A&A...311..230C} and ESPaDOnS periods was chosen so as to allow the possibility of large changes, whilst still restricting the window to a somewhat plausible range. 

Our analysis of the new reduction of the SMEI light curve, obtained between 2003 and 2011, finds $P_{\rm rot} = 0.97890(5)$~d from the full dataset. The analysis follows the same methods described by \citet{2016A&A...593A.106R}, but due to the relatively low data quality the wavelet analysis is hardly distinguishable from noise. The data were analysed in several subsets, by camera, by year, and over several years. All analyses gave identical results within uncertainty, although of course every one of these analyses provided larger errors than that obtained from the full dataset. Only the results for the analysis of the full data set are used here. 

The results in Fig.\ \ref{hd142990_bz_phot_rf_constp} are not improved by phasing the available data using any of the alternative periods. No matter which period is chosen, there are significant phase shifts introduced between photometric and magnetic datasets. In the case of the SMEI data, which must be binned by phase in order for a coherent variation to be easily discernable, the periods derived from the earlier photopolarimetric, Hipparcos, or Str\"omgren photometry lead to a much smaller amplitude in the phase-binned data, similarly to the results for the full magnetic dataset (middle panel of Fig.\ \ref{hd142990_bz_phot_rf_constp}).

\subsection{Photometric bandpass dependence}\label{bandpass}

   \begin{figure}
   \centering
   \includegraphics[trim=40 20 30 30, width=\hsize]{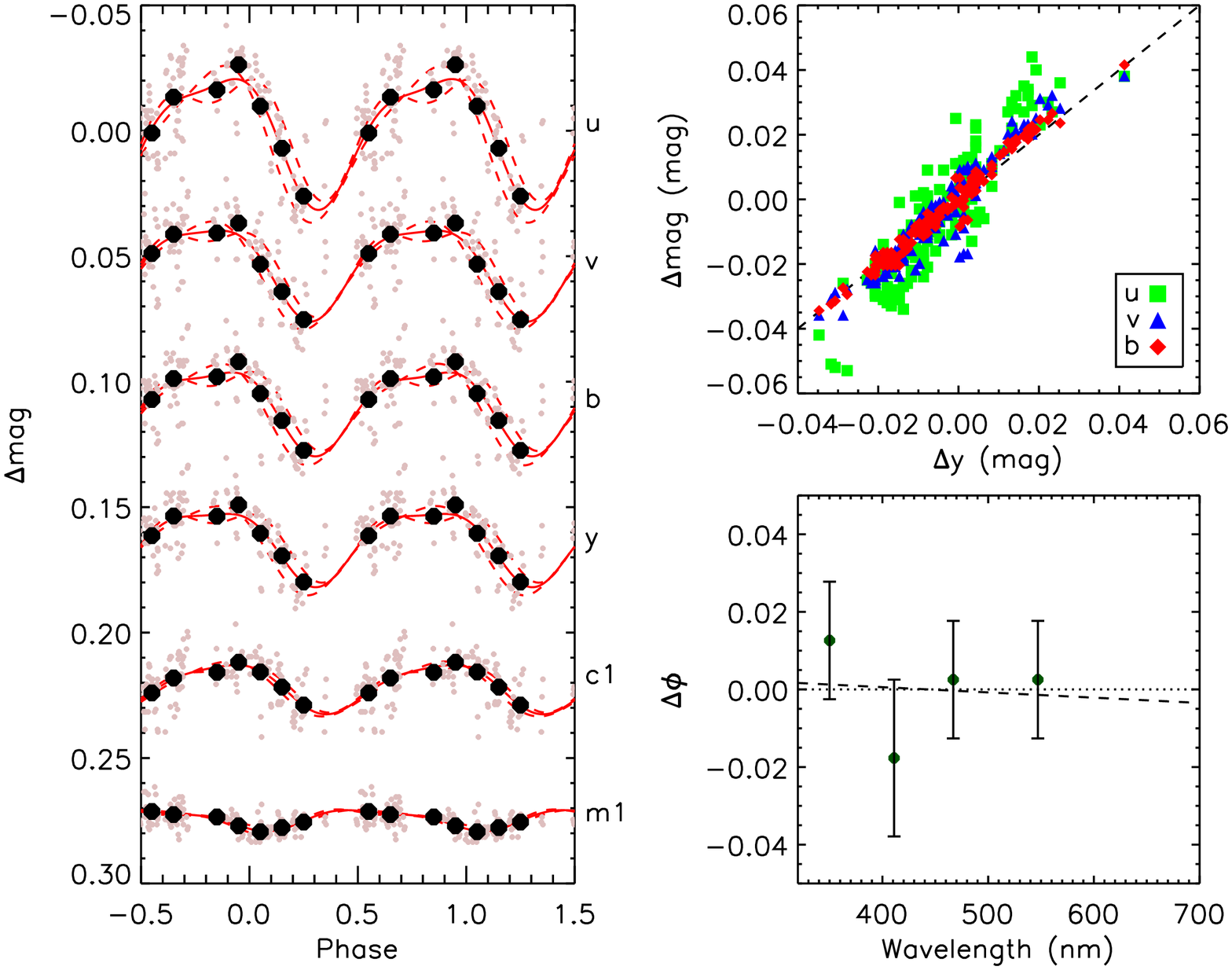} 
      \caption[]{Str\"omgren $uvby$ photometry and colour indices ({\em left}) from \protect\cite{1995A&AS..113...31S} phased with the $P_{\rm rot}$ inferred from the same data. The data have been vertically offset for display purposes. Individual measurements are shown by small grey circles, phased-binned measurements by large black circles. Solid and dashed curves show harmonic fits and 1$\sigma$ uncertainties. {\em Top right}: Str\"omgren $\Delta u$, $\Delta v$, and $\Delta b$ as functions of $\Delta y$. The dashed line shows $\Delta x=\Delta y$. {\em Bottom right}: the phase of minimum light as determined via the harmonic fits, as a function of the central wavelengths of the $uvby$ filters.}
         \label{hd142990_strom}
   \end{figure}

The photometric datasets were obtained using different filters with a variety of passband widths and central wavelengths. The light curve variations of CP stars are a consequence of chemical spots, which do not affect all regions of the spectrum in exactly the way; thus, there may be differences in the shapes of light curves obtained using different filters, which in some cases manifest as apparent phase shifts \citep[e.g.][]{2009A&A...499..567K,2012A&A...537A..14K,2015A&A...576A..82K}. We used the Str\"omgren $uvby$ photometry published by \cite{1995A&AS..113...31S} to evalute the degree to which HD\,142990's light curve morphology is affected by the choice of filter. Fig.\ \ref{hd142990_strom} shows the photometric magnitudes and colour indices phased with the rotation period inferred from these data. 

While there is some suggestion of a variation in $c_1$ with rotation phase, $m_1$ is almost constant. Variability in $c_1$ is likely due to changes in the vicinity of the Balmer jump, as reported for other stars by \cite{2009A&A...499..567K,2012A&A...537A..14K}. The lack of variation in $m_1$ suggests almost no difference in the behaviour of $v$, $b$, and $y$. The top right panel of Fig.\ \ref{hd142990_strom} shows $u$, $v$, and $b$ as functions of $y$, and verifies that while there are systematic differences between $u$ and $y$, $v$, $b$ and $y$ return almost identical results. We used harmonic fits of 2$^{nd}$ degree to $uvby$ (Fig.\ \ref{hd142990_strom}, left panel) to determine the phase of minimum light in each waveband, which is essentially constant for all 4 filters (Fig.\ \ref{hd142990_strom}, bottom right). 

Hipparcos, Kepler, and SMEI have central wavelengths of about 500, 600, and 700 nm respectively, and the central wavelength of the $y$ filter is about 550 nm; since there is essentially no phase shift between $y$ and the other Str\"omgren filters, it is almost certainly the case that the phase shifts in Fig.\ \ref{hd142990_bz_phot_rf_constp} --the smallest of which is about 0.1 cycles -- cannot be ascribed to the difference in filters. 

\subsection{Period evolution}

   \begin{figure}
   \centering
   \includegraphics[trim=40 40 20 0, width=\hsize]{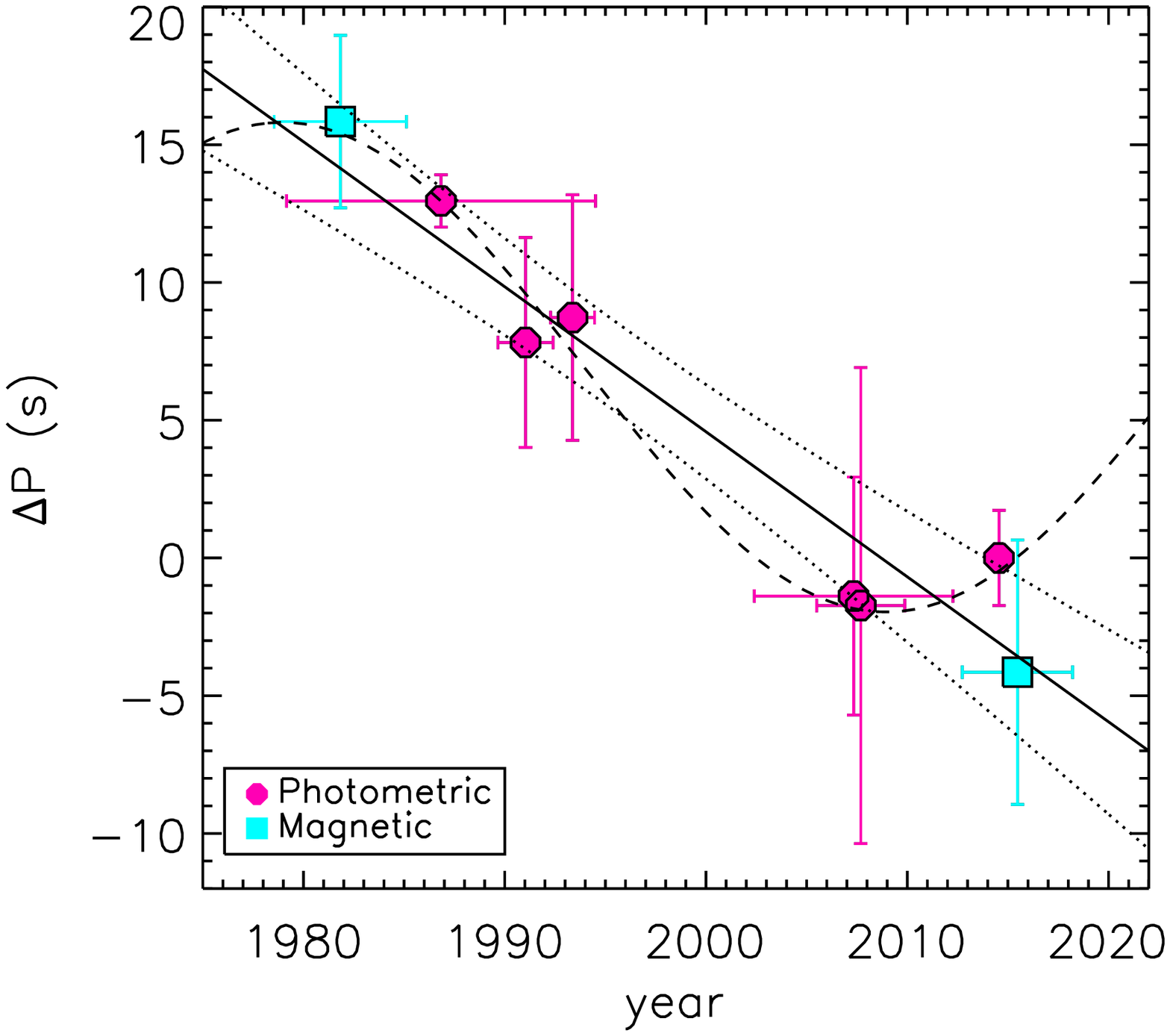} 
      \caption[]{Evolution of the rotational period of HD\,142990. The x-axis positions correspond to the mean HJD of the dataset from which the period was determined, and horizontal error bars indicate the time span of the dataset. In temporal sequence, the photometric periods were obtained via Str\"omgren $y$ \protect\citep{1996A&A...311..230C}, Hipparcos, Str\"omgren $y$ \protect\citep[using data from ]{1995A&AS..113...31S}, SMEI, STEREO \protect\citep{2012MNRAS.420..757W}, and K2 \protect\citep{2018A&A...616A..77B}. The two periods determined from \bz~measurements correspond to photopolarimetric data and ESPaDOnS data. The solid and dotted lines show the least-squares linear fit and uncertainties. The dashed curve shows the least-squares sinusoidal fit assuming an oscillatory period change with a 60-year period.}
         \label{hd142990_deltap}
   \end{figure}

   \begin{figure}
   \centering
   \includegraphics[trim=40 40 20 0, width=\hsize]{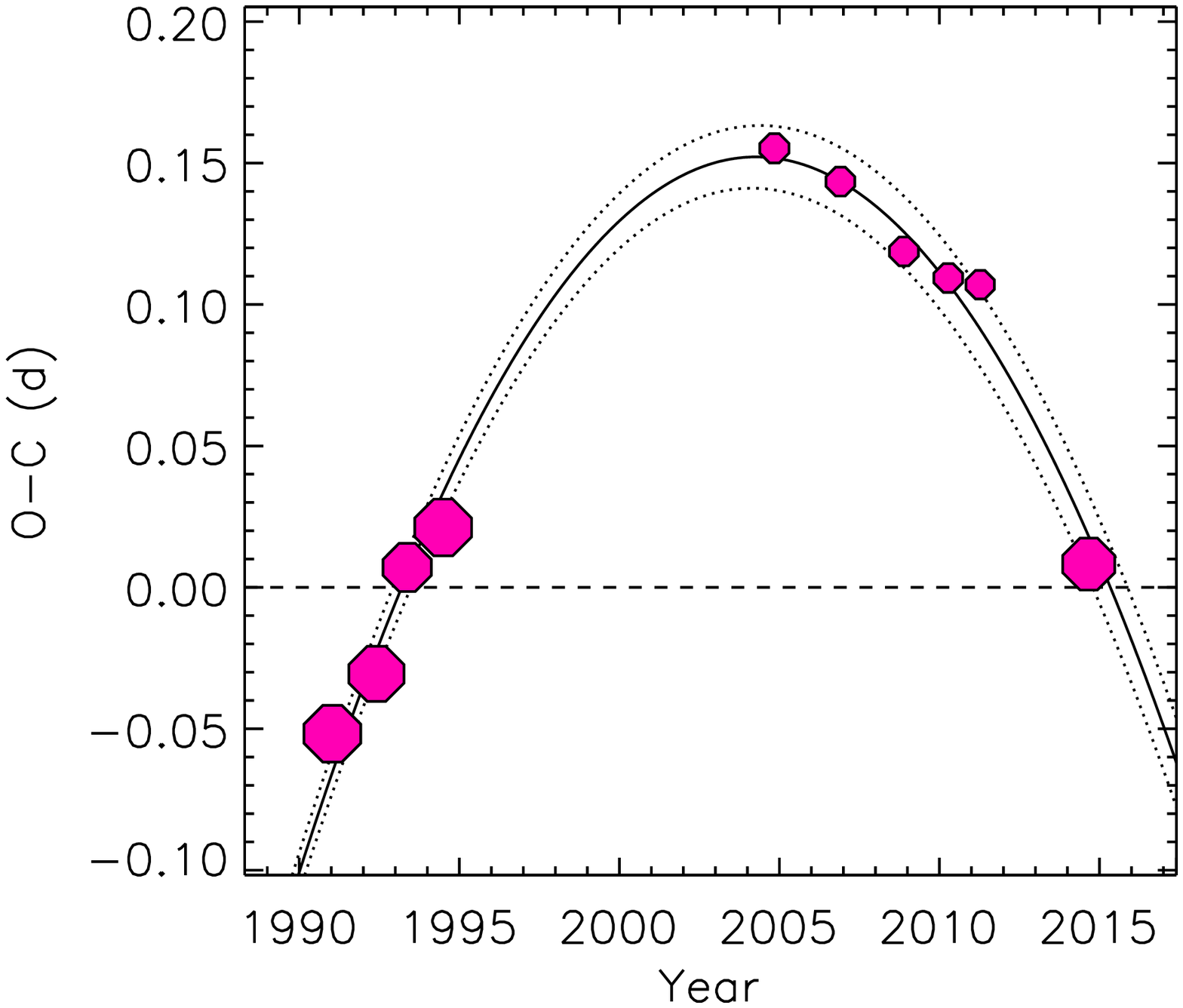} 
      \caption[]{$O-C$ diagram for the photometric datasets of sufficient size to fully sample the phase diagram (K2, SMEI, \protect\cite{1995A&AS..113...31S} Str\"omgren $y$, and Hipparcos). Symbol size is proportional to the inverse of the mean phase-binned variance. The solid and dotted curves show a parabolic fit and the fit uncertainties.}
         \label{hd142990_oc}
   \end{figure}

\begin{table}
\caption[]{Summary of $O-C$ values for individual 4-year time bins. Year and HJD give the mean for each bin; Dataset gives the origin of the data corresponding to the reference key in Table \ref{obstab}; Variance gives the mean phase-binned variance; $N_{\rm obs}$ gives the number of observations in each time bin.}
\label{octab}
\begin{tabular}{l l l l l l}
\hline\hline
Year & HJD - & Dataset & $O-C$ & Variance & $N_{\rm obs}$\\
     & 2400000 (d) &         & (d)   & (mag)  &  \\
\hline
1991 & 48307 & $e$ &  -0.062  &    0.000024 & 111  \\
1992 & 48809 & $e$ & -0.040   &   0.000007 & 8 \\
1993 & 49160 & $g$ &  0.000   &   0.000067 & 142 \\
1994 & 49571 & $g$ &  0.011   &   0.000014 & 21 \\
2004 & 53348 & $h$ &  0.149   &   0.003356 & 8465 \\
2006 & 54101 & $h$ &  0.139   &   0.003570 & 8224 \\
2008 & 54825 & $h$ &  0.112   &   0.003574 & 8590 \\
2010 & 55330 & $h$ &  0.101   &   0.003596 & 6178 \\
2011 & 55694 & $h$ &  0.099   &   0.003887 & 1796 \\
2014 & 56932 & $j$ &  0.002   &   0.000044 & 3293 \\
\hline\hline
\end{tabular}
\end{table}

Fig.\ \ref{hd142990_deltap} shows the inferred change in period as a function of time, where we chose $\Delta P = 0$ as the K2 period. $\Delta P$ is given in Table \ref{prottab}, where the periods selected for analysis are indicated in boldface. With the exception of the \cite{2011MNRAS.414.2602D} period (which is simply a duplicate of the value we determined ourselves from Hipparcos photometry), periods were rejected either because either the precision was too low \citep[e.g. the period determined by][]{1977A&AS...30...11P}, or the determinations were judged to be inaccurate, as explained above in \S~\ref{subsec:published_periods} and \ref{subsec:period_analysis}. Inclusion of the \cite{2011MNRAS.414.2602D} or \cite{1977A&AS...30...11P} periods has no effect on results. Inclusion of the other periods, many of which are nominally highly precise, makes any pattern difficult to discern due to scatter of up to 100s of s as compared to the K2 period. 

All of the modern periods -- those determined from datasets obtained since 2000 -- are consistent with one another, and are about 20 seconds {\em shorter} than the periods determined in the 1980s and 90s (which are also consistent with one another). The period change between the 1980s and the 2010s is well-matched by a linear decrease of $-0.53 \pm 0.12$~s/yr (solid line in Fig.\ \ref{hd142990_deltap}). 

To verify that the period is changing in a coherent fashion, we constructed an $O-C$ (Observed minus Calculated) diagram (Fig.\ \ref{hd142990_oc}). Only those photometric datasets of sufficient size and quality to fully sample the phase diagram were used, i.e.\ the K2, SMEI, Hipparcos, and \cite{1995A&AS..113...31S} Str\"omgren $y$ data. We phased the K2 data with the K2 period, binned the data by phase, and fit a second-order sinusoid. We then broke the remaining data into time segments, phase-binned the data in each time segment, and determined the phase shift relative to the K2 photometry that was required to minimize the reduced $\chi^2$ of the harmonic fit. In Fig.\ \ref{hd142990_oc} we show results for 4-year time segments; the results do not change qualitatively for different time segment durations. Results are also tabulated in Table \ref{octab}. 

   \begin{figure}
   \centering
   \includegraphics[trim=30 30 30 30, width=8.5cm]{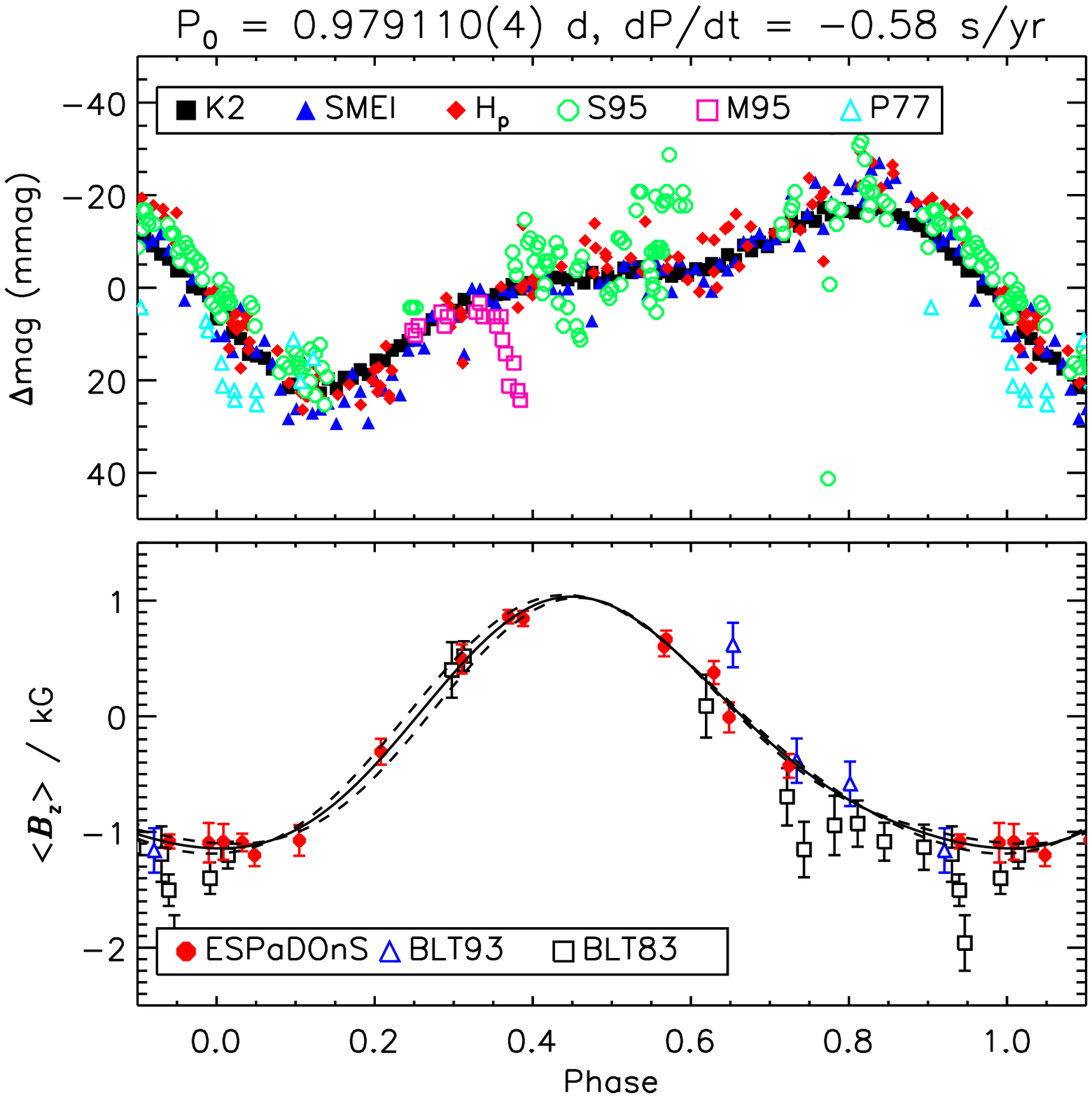} 
      \caption[]{As Fig.\ \protect\ref{hd142990_bz_phot_rf_constp}, but phased with the variable ephemeris described in the text.}
         \label{hd142990_bz_phot_rf_changingp}
   \end{figure}


The parabolic shape of the $O-C$ curve is a clear indication of a changing period, since the phase shift between datasets that would be produced by a constant period would be a straight line. Obtaining $\dot{P}$ from the quadratic coefficent of the parabolic fit to the $O-C$ diagram \citep[e.g.][]{2005ASPC..335....3S} yields $\dot{P}=-0.58 \pm 0.01$~s/yr, which is compatible with the rate of period change inferred from the linear fit to $\Delta P$. 

The phase $\phi$ of the variable ephemeris is given by

\begin{equation}\label{phi}
\phi(t) = \frac{t - T_N}{P_N}~{\rm mod}~1,
\end{equation}

\noindent where $t$ is in HJD, and $P_N$ and $T_N$ are the period and zero-point at cycle $N$:

\begin{align}\label{pntn}
P_N &= P_0 + \dot{P} N P_0 \\
T_N &= T_0 + P_0 N + \frac{\dot{P} N^2 P_0}{2},
\end{align}

\noindent where $\dot{P}$ is in units of d/d, and $N$ is found by

\begin{equation}\label{cyclen}
N = \frac{2\Delta t}{2 P_0 + \dot{P} \Delta t},
\end{equation}

\noindent with $\Delta t = t - T_0$, where it is assumed that $\dot{P}/P_0\Delta t \ll 1$. 

Fig.\ \ref{hd142990_bz_phot_rf_changingp} shows the various datasets phased using Eqns.\ \ref{phi}--\ref{cyclen}, using $P_0 = 0.979110(4)$~d, $T0 = 2442820.93(3)$, and $\dot{P} = -0.58 \pm 0.01$~s/yr. In contrast to the various constant periods examined in Fig.\ \ref{hd142990_bz_phot_rf_constp}, there are no obvious phase shifts between comparable datasets. These values and uncertainties were obtained by two methods. First, starting from the ESPaDOnS period and the ESPaDOnS epoch obtained from a second-order harmonic fit to the ESPaDOnS data, we solved Eqn.\ \ref{pntn} for $T_0$ and $P_0$ at the time of the \cite{1977A&AS...30...11P} photometry using $\dot{P}=-0.58 \pm 0.02$ s/yr. By varying $\dot{P}$, we found that values in the range of $-0.58 \pm 0.01$ s/yr are able to phase the magnetic and photometric datasets without introducing phase shifts larger than the scatter in the data. Our second method was to use phase dispersion minimization, starting with the K2 period, taking the epoch as the mean HJD of the K2 dataset, and again letting $\dot{P}$ vary within $-0.58 \pm 0.02$ s/yr. The result of this test was that the minimum variance was obtained with a period of $0.97885(2)$ d at the time of the K2 dataset -- which is consistent with the ESPaDOnS period, but not with the published K2 period -- and yielded $P_0 = 0.97911(2)$ at the time of the \cite{1977A&AS...30...11P} photometry (which is consistent with the results obtained from the first method). The slight inconsistency with the period obtained from the K2 data using Lomb-Scargle analysis can likely be reconciled if either the PDM or Lomb-Scargle uncertainty is slightly under-estimated, as it is only a 3$\sigma$ difference w.r.t.\ the uncertainties.

\section{Discussion \& Conclusions}\label{sec:conclusion}

   \begin{figure*}
   \centering
\begin{tabular}{cc}
   \includegraphics[trim=30 30 30 30, width=0.5\textwidth]{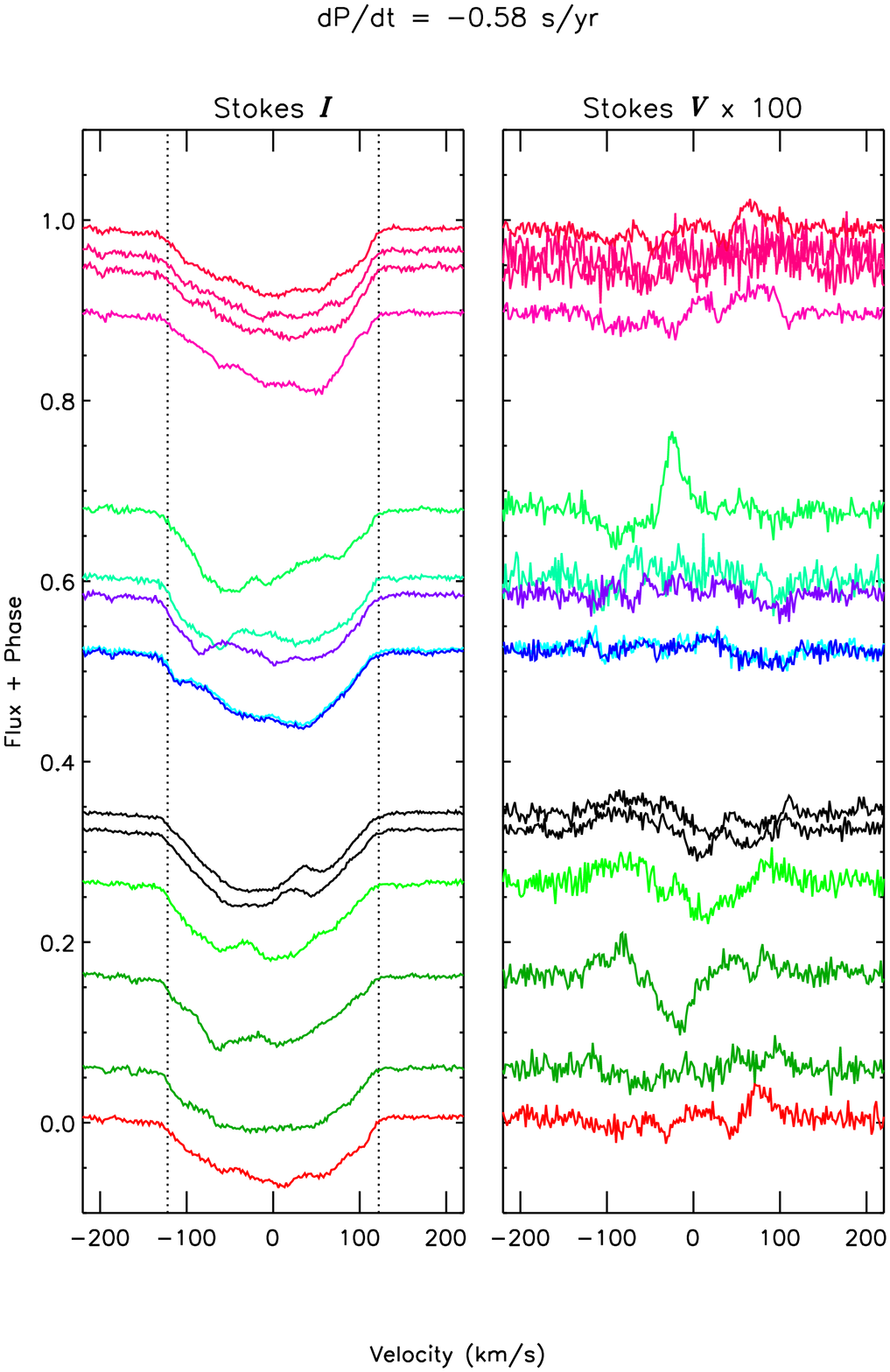} 
   \includegraphics[trim=30 30 30 30, width=0.5\textwidth]{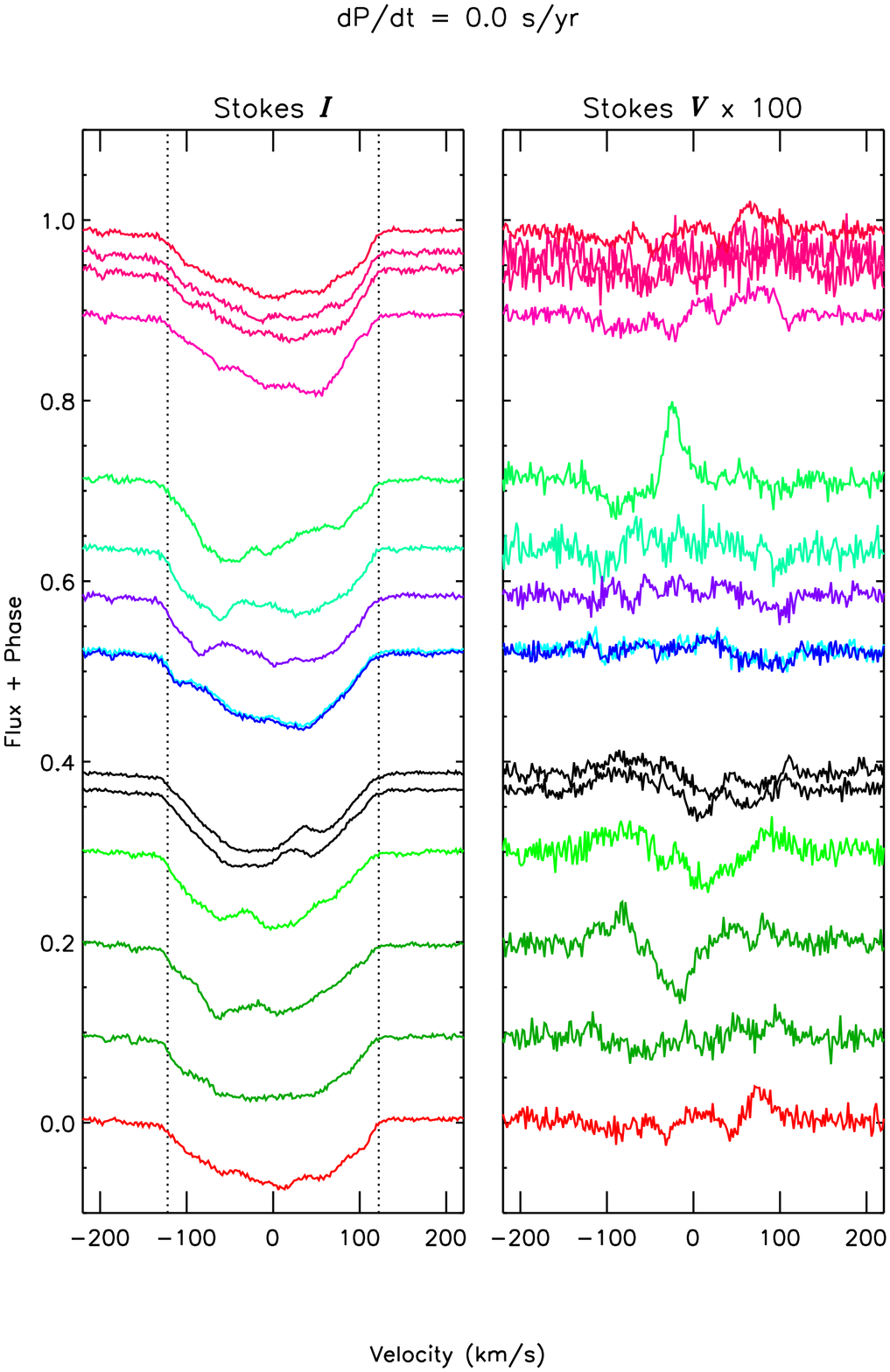} 
\end{tabular}
      \caption[]{Stokes $I$ (left sub-panels) and Stokes $V$ (right sub-panels) LSD profiles phased with the variable ephemeris (left panels) and the period determined from ESPaDOnS data (right panels). Colour indicates cycle number. Vertical dotted lines indicate $\pm$\vsini.}
         \label{s18_met_lsd_phased}
   \end{figure*}

We have shown that 1) a constant period cannot coherently phase the available photometric and magnetic data; 2) photometric phase shifts that can be plausibly attributed to differences in the various bandpasses are much smaller than those obtained between the various photometric datsets using constant periods; 3) periods obtained independently from individual datasets show a coherent decrease over time; 4) the photometric $O-C$ diagram is also consistent with an accelerating rotation period; and finally 5) phasing the data with a rotational period that accelerates at the rate of $-0.58 \pm 0.02$ s/yr is able to coherently phase the data. It is worth emphasizing that the magnetic data, which was acquired over a 35-year period, is coherently phased by the variable ephemeris derived from the photometric, representing an unbiased test of the photometric $O-C$ results. 

One possible, conventional, explanation for an apparently accelerating rotational period may be the light-time effect due to the orbit of a binary companion. However, none of the magnetic B-type stars in which this phenomenon has been detected, including the present star, are known to be in binary systems (\cite{2002A&A...382...92S} conducted a NIR search for visual companions, and found no evidence of a companion in the case of HD\,142990). The change in period $\Delta P$ due to the light-time effect should correspond to a change in radial velocity $\Delta {\rm RV} = c \Delta P / P$, where $c$ is speed of light \citep[e.g.][]{1992A&A...253..178P}. Fig.\ \ref{s18_met_lsd_phased} shows the least-squares deconvolution (LSD) profiles extracted from the ESPaDOnS dataset with a line mask using all metallic lines in the spectrum \citep[for details see][]{2018MNRAS.475.5144S}. No bulk RV variability is apparent. Measuring the RV is complicated by the spectroscopic variability introduced by chemical spots, which in addition to equivalent width changes also introduce RV variations coherent with the rotation phase due to changes in the star's centre of gravity. As a result, only measurements performed on observations obtained close to the same rotation phase can be compared. The ESPaDOnS data contains two observations obtained close to phase 0.6, separated by about 3 years (one on 14/06/2014, the second on 14/05/2017, with a difference in phase of 0.02 cycles when phased using the variable ephemeris). Measuring the centres-of-gravity of the LSD profiles extracted from these observations yields a difference in RV of 1~\kms, comparable to the measurement uncertainty. The RV change expected over 3 years if $\dot{P}$ is due to orbital motion is about 3 \kms, so this test must be considered inconclusive. However, a change of $-20$~s over the 30 years of observations should have led to $\Delta {\rm RV} = 71$~\kms; it is unlikely that such a large change in RV would have been missed. The Pulkovo Compilation of Radial Velocities \citep{2006AstL...32..759G} give RV$=-12 \pm 3$~\kms, consistent with RVs measured from ESPaDOnS data (which have a mean and standard deviation of $-4$ and $5$~\kms), suggesting that the RV has been stable over a time span of at least a decade.

Another explanation may be that HD\,142990 is still evolving towards the zero-age main sequence, and that rotational spin-up is a consequence of ongoing core-contraction. The star is a member of the Upper Sco OB association \citep{1999AJ....117..354D}, which has an estimated age of $\log{(t/{\rm yr})} = 6.7 \pm 0.1$ \citep{land2007}. Given the star's mass \citep[about 5 \msun;][]{land2007}, it is indeed very close to the ZAMS. The possibiity that its core might still be contracting should be explored, once grids of evolutionary models for OB stars with surface fossil magnetic fields become available \citep[e.g.][]{2019MNRAS.485.5843K}.

\cite{2018CoSka..48..203M} suggested that vertically stratified differential rotation, due to episodic magnetic coupling and decoupling of the upper and lower layers of the photosphere, may explain the phenomenon for CU\,Vir and HD\,37776. In this scenario, when the upper and lower layers couple, angular momentum is transported to the upper layer, spinning it up; when they decouple, the outer layer sheds angular momentum via magnetic braking. An alternate mechanism was proposed by \cite{2017MNRAS.464..933K}, who suggested torsional oscillations arising from magnetohydrodynamic waves. However, they noted that while this mechanism can explain the oscillatory period of CU Vir, it cannot explain the behaviour of HD\,37776. It may be interesting to see if this hypothesis is plausible in the case of HD\,142990.

Line profile variations are in principle a sensitive diagnostic of rotational phase \citep[e.g.][]{2017A&A...605A..13K}. Fig.\ \ref{s18_met_lsd_phased} compares the phasing of the ESPaDOnS LSD profiles obtained by the variable ephemeris and the period inferred from ESPaDOnS data. Different rotational cycles are indicated with different colours. In most cases, observations with similar phases were obtained at similar times, and so are insensitive to period evolution. Observations obtained at different rotational cycles, but with similar computed phases, can be seen near phases 0.5 and 0.6. The former are almost identical in phase, but were obtained only 1 rotational cycle apart; unsurprisingly, Stokes $I$ and $V$ are almost indistinguishable. Near phase 0.6, the observations differ by about 0.02 cycles with the variable ephemeris and 0.05 cycles with the constant ephemeris, and are separated by 1065 d. Stokes $I$ and $V$ are both similar between these observations; however, the morphological change seems too fast with the variable ephemeris, while phasing the line profiles with the ESPaDOnS period seems to give a somewhat improved phasing of these two observations. The relatively small size of the dataset and small number of observations overlapping in phase makes this qualitative test inconclusive, but suggestive. 

While we have assumed a constant acceleration of the period, there is no reason to believe this must be the case. In fact, te other two stars in which rotational acceleration has been reported exhibit apparently cyclical changes in $P_{\rm rot}$ and $\dot{P}$ \citep{miku2011,2017ASPC..510..220M}. The top panel of Fig.\ \ref{hd142990_deltap} shows a sinusoidal fit to $\Delta P$, where we arbitarily assumed a 60-year periodicity (or about twice the current span of observations). Notably, while $P_{\rm rot}$ has apparently changed by about 20 s between 1980 and 2010, between 2005 and 2015 the results are consistent with no change in period. A cyclic variation in $\dot{P}$ could explain why the phasing of the ESPaDOnS data is improved by a constant ephemeris. Further photometric monitoring will be essential to distinguishing between these scenarios. If the suggestion by \cite{2018MNRAS.478.2835L} that HD\,142990 exhibits pulsed radio emission is confirmed, this phenomenon may also enable tight constraints on $\dot{P}$ \citep[e.g.][]{miku2011}.

It is interesting to note that the measured period change of HD\,142990, about 20 s, is similar to the lower limit of that of HD\,37776 (although it is likely that the amplitude of HD\,37776's period change is much higher), and much greater than that of CU Vir (about 4 s). CU Vir is a more rapid rotator ($P_{\rm rot} \sim 0.52$~d) than either HD\,142990 or HD\,37776~($P_{\rm rot} \sim 1.5$~d); likewise, HD\,142990 is intermediate in mass between CU Vir (a late Bp star) and HD\,37776 (a hot He-strong B2 star). CU Vir and HD\,37776 have both been mapped via Zeeman Doppler Imaging; the former possesses a distorted dipolar magnetic field topology with a mean surface strength of about 4~kG \citep{2014A&A...565A..83K}, while the latter has an extremely complex topology with a maximum local magnetic field modulus of around 30 kG \citep{koch2011}. As can be seen in Fig.\ \ref{s18_met_lsd_phased}, the phase curve is not yet sampled with sufficient density to perform Zeeman Doppler Imaging; however, HD\,142990's anharmonic \bz~curve shows signs of departure from a purely dipolar magnetic field, so we can infer that its surface magnetic field is likely to be qualitatively similar to that of CU Vir in both topology and strength. HD\,142990 is intermediate between CU Vir and HD\,37776 in stellar and rotational properties, and likely similar to CU Vir in magnetic properties. Assuming a common mechanism, some or all of these factors may explain why its period change is apparently intermediate in amplitude between CU Vir and HD\,37776. 

The remarkable occurrence of rotational spin-up in 3 of the 4 stars (CU\,Vir, HD\,37776, $\sigma$ Ori E, and now HD\,142990) for which period change has been directly measured suggests that this may well be a general phenomenon. $\sigma$ Ori E -- the only exception so far -- should be monitored in the future for signs of rotational acceleration. If the phenomenon is indeed common, this suggests a new element in our understanding of the rotational evolution of magnetic, hot stars. Given that magnetic stars are known to be much more slowly rotating than non-magnetic stars as a population, magnetic braking must dominate over the long term. However, superimposed on this long-term trend may be an oscillatory pattern of spin-up and spin-down, driven by entirely different physics. This may complicate efforts to compare theoretical spin-down timescales to observations (since in this case multiple period oscillation cycles would need to be observed, each likely to be decades in length). On the other hand, the phenomenon may provide otherwise unobtainable insights into the internal structure and evolution of magnetic hot stars. 


\section*{Acknowledgements}

This work is based on observations obtained at the Canada-France-Hawaii Telescope (CFHT) which is operated by the National Research Council of Canada, the Institut National des Sciences de l'Univers of the Centre National de la Recherche Scientifique of France, and the University of Hawaii. This research has made use of the VizieR catalogue access tool, CDS, Strasbourg, France. The original description of the VizieR service was published in \cite{2000A&AS..143...23O}. Some of the data presented in this paper were obtained from the Mikulski Archive for Space Telescopes (MAST). STScI is operated by the Association of Universities for Research in Astronomy, Inc., under NASA contract NAS5-26555.  This paper includes data collected by the Kepler mission. Funding for the Kepler mission is provided by the NASA Science Mission directorate. MS acknowledges support fom the Annie Jump Cannon Fellowship, supported by the University of Delaware and endowed by the Mount Cuba Astronomical Observatory. GAW acknowledges support from a Discovery Grant from NSERC. PC acknowledges support from the Department of Science and Technology via SwarnaJayanti Fellowship awards (DST/SJF/PSA-01/2014-15). The authors thank Andrzej Pigulski for evaluating the ASAS-SN and SuperWASP databases. 

\bibliography{bib_dat.bib}{}


\end{document}